\documentstyle[11pt,paspconf,epsf]{article}
\def\edcomment#1{\iffalse\marginpar{\raggedright\sl#1\/}\else\relax\fi}
\reversemarginpar

\def\lsim{\lower.5ex\hbox{$\; \buildrel < \over \sim \;$}}
\def\gsim{\lower.5ex\hbox{$\; \buildrel > \over \sim \;$}}

\begin{document}
\title{Nonthermal X-Rays from the Galactic Ridge: a Tracer of Low Energy
Cosmic Rays ?}

\author{V. Tatischeff}
\affil{Centre de Spectrom\'etrie Nucl\'eaire et de Spectrom\'etrie de
Masse, IN2P3-CNRS, 91405 Orsay, France}

\author{R. Ramaty and A. Valinia}
\affil{Laboratory for High Energy Astrophysics, Goddard Space Flight Center,
Greenbelt, MD 20771}


\begin{abstract}

A distinct low energy cosmic-ray component has been proposed to
explain the essentially constant Be/Fe ratio at low metallicities.
Atomic collisions of such low energy ions produce characteristic
nonthermal X-ray emission. In this paper, we study the possible
contribution of such X-rays to the Galactic ridge emission. We
show that they would account for $\lsim$10\% of the 10-60 keV
luminosity of the thin Galactic disk component detected with {\it
RXTE}. They could make a more significant contribution in the
0.5-10 keV energy range, provided that the nonthermal ion
population extends down to about 1 MeV/nucleon and delivers about
10$^{42}$ erg s$^{-1}$ to the interstellar medium, comparable to
the total power suplied by the Galactic supernovae. But since the
nonthermal X-rays in this energy range are essentially produced
below the thresholds of the Be-producing cross sections, their
detection does not necessarily imply a low energy cosmic-ray
origin for the spallogenic light elements. A significant
contribution of nonthermal X-rays could alleviate the problem of
the origin of the hard component observed with {\it ASCA} in the
Scutum arm region.

\end{abstract}

\section{Introduction}

The recent measurements of Be and B abundances in low metallicity
stars (see Vangioni-Flam et al. 1998 for a recent compilation)
shed new light on the origin of cosmic rays (Ramaty, Kozlovsky, \&
Lingenfelter 1998). A distinct Galaxy wide cosmic-ray component,
accelerated out of fresh nucleosynthetic matter and predominant at
low energies ($\lsim$100 MeV/nucleon), has been proposed to
account for the quasi linear correlation between Be and Fe for
\mbox{[Fe/H]$<$-1} (Cass\'e, Lehoucq \& Vangioni-Flam 1995;
Ramaty, Kozlovsky \& Lingenfelter 1996). Alternatively, it was
suggested that the standard Galactic cosmic rays themselves are
accelerated mostly out of supernova ejecta (Lingenfelter, Ramaty,
\& Kozlovsky 1998; Higdon, Lingenfelter, \& Ramaty 1998). Nuclear
gamma-ray line observations could distinguish between the models.
Low energy cosmic rays (LECRs) also produce X-rays by a variety of
processes (Tatischeff, Ramaty, \& Kozlovsky 1998 and references
therein; Dogiel et al. 1998). Electron capture and excitation in
low energy ions (typically in the energy range from a few tenths
to several tens of MeV/nucleon) produce characteristic broad X-ray
lines in the 0.5-10 keV energy range and K-shell vacancy creation
in ambient heavy atoms by similar energy fast ions produce narrow
X-ray lines also up to about 10 keV. In addition, ions of energies
up to a few hundreds of MeV/nucleon produce X-ray continuum at
energies of several tens of keV. X-ray observations, therefore,
can provide further information, and also set constraints, on
LECRs in the Galaxy.

The diffuse X-ray emission from the Galactic disk has been
intensively studied since its first detection by Bleach et al.
(1972). Yamauchi et al. (1996) and Kaneda et al. (1997) have argued,
from deep {\it ASCA} observations of the Scutum arm region, that the
bulk of the Galactic ridge X-ray emission (GRXE) is truly diffuse,
as opposed to resulting from the superposition of unresolved point
sources. The {\it ASCA} spectrum shows K lines from Mg, Si, S and Fe
(at energies from about 1 to 9 keV), and has been modeled by a
double-temperature non-equilibrium ionization plasma model with
temperatures of kT$\sim$0.8 keV and kT$\sim$7 keV (Kaneda et al.
1997). The lower temperature emission could be associated with a
population of supernova remnants and superbubbles in the Galactic
disk (Kaneda et al. 1997; Valinia \& Marshall 1998). The origin of
the component at the higher plasma temperature, hereafter the hard
component, is more problematic. In particular, it is difficult to
explain how a $\sim$7 keV plasma could be confined to a very low
scale height (b$\sim$0.5$\deg$), as its temperature significantly
exceeds the gravitational escape temperature ($\sim$0.4 keV) of the
Galaxy.

The GRXE above 10 keV is clearly of nonthermal origin. A power law
tail reaching 600 keV has been detected with {\it Ginga} and
Welcome-1 (Yamasaki et al. 1997) as well as with OSSE (Skibo et
al. 1997). Valinia \& Marshall (1998) have recently performed a
careful scan of the Galactic plane with RXTE. They extract two
spatial components from the GRXE in the 2-60 keV energy range: a
thin disk of width $\lsim$0.5$\deg$ and a broad component whose
latitude distribution is approximated by a Gaussian of
$\sim$4$\deg$ FWHM. They discuss the origin of the $>$10 keV
X-rays in the two spatial components in terms of unresolved
discrete sources, inverse Compton scattering and bremsstrahlung of
fast electrons, as well as inverse bremsstrahlung from energetic
protons.

In this paper we present calculations that demonstrate the
relationship between the Be production and the X-ray and nuclear
gamma-ray line productions, and we investigate the possible
contribution of nonthermal ion interactions to the GRXE. We use the
constraint set by the essentially constant Be/Fe ratio at low
metallicities on the current epoch Be production (Ramaty et al.
1997) to calculate the expected X-ray luminosity from nonthermal ion
interactions. We compare it with the observed luminosities of the
two spatial components detected with RXTE and with the luminosity of
the hard component in the {\it ASCA} energy band. We also calculate
the expected 3-7 MeV nuclear gamma-ray line flux from the central
radian of the Galaxy. In agreement with previous calculations
(Ramaty et al. 1997), we show that this emission would be only
marginally detectable with {\it INTEGRAL}.

\section{Interaction model}

We perform the calculations in a steady state, thick target model
in which accelerated particles, injected at a constant rate,
interact with a neutral ambient medium of solar composition. The
gamma-ray line, Be and X-ray productions are calculated as in
Ramaty et al. (1996; 1997) and Tatischeff et al. (1998),
respectively. We employ three different compositions for the fast
ions: CRS and CRS$_{\rm metal}$ (Ramaty et al. 1997) and OB$_{\rm
IG}$ (Parizot, Cass\'e \& Vangioni-Flam 1997, table 1, the OB/0.04
column). CRS is the composition of the current epoch Galactic
cosmic-ray sources; CRS$_{\rm metal}$ is identical to CRS, but
without protons and $\alpha$ particles. OB$_{\rm IG}$ is the
calculated average composition of the stellar winds from OB
associations in the inner Galaxy; the O-to-proton ratio for this
composition is similar to the corresponding ratio for the CRS
composition, but the abundances of Mg, Si, S and Fe relative to
protons are lower.

\begin{figure}
\plotone{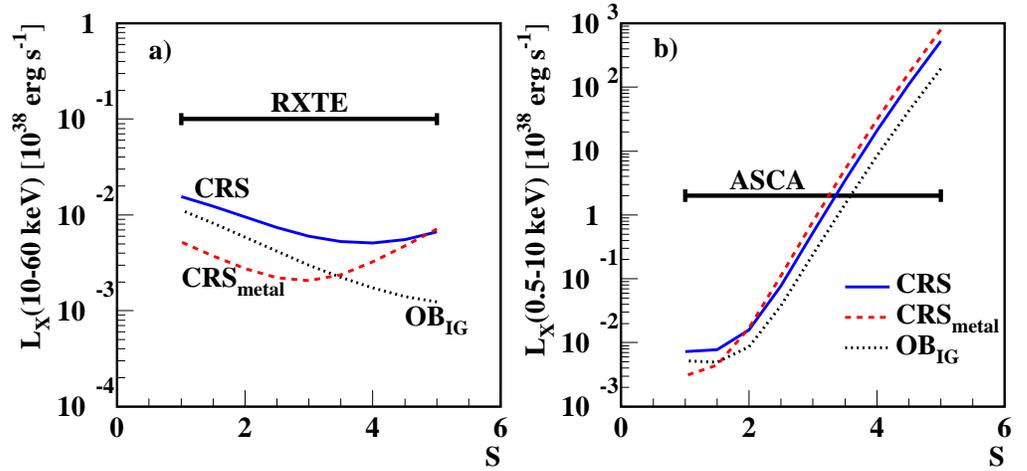} \caption{Predicted X-ray luminosities (a) in
the 10-60 keV energy range and (b) in the 0.5-10 keV energy range,
as a function of power-law spectral index (eq.~1). The
calculations assume that all the current era Be production is due
to LECR interactions, with a Be production rate of
3$\times$10$^{39}$ atoms s$^{-1}$. Also shown are the estimated
10-60 keV luminosity of the thin disk component detected with RXTE
(panel a) and the luminosity of hard component in the 0.5-10 keV
band from the latest {\it ASCA} survey (panel b). Photoelectric
absorption has been removed in the {\it ASCA} analysis and is thus
not taken into account in the calculations.}
\end{figure}

As the X-ray line production is sensitive to very low energy ions
(near 1 MeV/nucleon), we replace the previously used ion
spectra (Ramaty et al. 1996), which were quite flat at low energies,
with an accelerated ion source spectrum given by a power law in
kinetic energy:
\begin{equation}
q_i(E) \propto E^{-s} {\rm ~~~for~1~MeV/nucleon<}E{\rm <10^3~MeV/nucleon.}
\end{equation}
The corresponding power is $\dot{W} = \sum_{i} A_i \int E q_i(E) dE$,
where $A_i$ is the nuclear mass for particle species $i$.

We normalize both X-ray and gamma-ray line productions to the
instantaneous Galaxy-wide Be production rate (Ramaty et al. 1997)
\begin{equation}
\dot{Q}(Be) = {Be \over Fe} \times {M_{SNII}(Fe) \over {56 m_p}}
\times \dot{SN} = 3\times10^{39} {\rm ~atoms~s^{-1}}.
\end{equation}
Here, $Be/Fe$=1.45$\times$10$^{-6}$ is the best fit constant value
(Vangioni-Flam et al. 1998) to the observed abundance ratio for
[Fe/H]$<$-1; $M_{SNII}(Fe)$=0.1 M$_{\odot}$ is the average Fe
production yield per core-collapse supernova (Ramaty \&
Lingenfelter 1998); $m_p$ is the proton mass; and $\dot{SN}$=3 per
century is the current epoch Galactic supernova rate.

\section{Results}

The calculated nonthermal X-ray luminosities are shown in Figure
1. In the 10-60 keV energy range (Fig. 1a), the bulk of the
emission is due to inverse bremsstrahlung (Tatischeff et al. 1998)
and radiative electron capture (REC) on fast Fe (Tatischeff \&
Ramaty 1999). REC is the dominant emission process for the CRS
composition and $s$$>$4.3, and for the CRS$_{\rm metal}$
composition and $s$$>$3.2. It is less important for the OB$_{\rm
IG}$ composition which, as just mentioned, is more impoverished in
Fe. As both Be-atoms and 10-60 keV X-rays are essentially produced
by ions of the same energies, the calculated luminosities are not
very dependent on the source spectrum. They range from $5.1 \times
10^{35}$ to $1.5 \times 10^{36}$ erg s$^{-1}$ for the CRS
composition, $2.1 \times 10^{35}$ to $7.1 \times 10^{35}$ erg
s$^{-1}$ for CRS$_{\rm metal}$ composition, and $1.2 \times
10^{35}$ to $1.1 \times 10^{36}$ erg s$^{-1}$ for OB$_{\rm IG}$
composition (Fig. 1a). Valinia \& Marshall (1998) estimated the
luminosities of the GRXE in the 10-60 keV band to be
1.5$\times$10$^{38}$ erg s$^{-1}$ and 10$^{37}$ erg s$^{-1}$ for
the broad component and the thin disk, respectively. We thus
conclude that nonthermal X-ray production from fast ion
interactions is probably not the main Galactic ridge emission in
the 10-60 keV energy range, unless the X-rays are produced in
source regions in which the Be is either destroyed or prevented
from escaping to the interstellar medium.

The calculated X-ray luminosities in the 0.5-10 keV energy range are rapidly
increasing functions of the source spectrum index (Fig. 1b). Indeed,
the X-rays in this energy range are essentially produced at low energies,
below the thresholds of the Be-producing spallation cross sections. We see
that for 3$<$$s$$<$3.5, nonthermal X-rays could account for most of the
hard component detected with {\it ASCA} in the 0.5-10 keV energy range
(Kaneda et al. 1997). We note, however, that the estimated luminosity of
this component, 2$\times$10$^{38}$ erg s$^{-1}$ (Kaneda et al. 1997), is
model dependent. It has been evaluated under the assumption of a very hot
plasma origin, and after removing the photoelectric absorption in the
{\it ASCA} data analysis. As a significant fraction of this luminosity is
contained in low energy X-rays, which are not detected because they are
completely absorbed in the Galactic plane, its value could be different
if part of the emission is of nonthermal origin.

Figure 2 shows the LECR power deposition into the interstellar
medium that accompanies the production of 2$\times$10$^{38}$ erg
s$^{-1}$ in the 0.5-10 keV energy range, the estimated X-ray
luminosity of the hard component observed with {\it ASCA}. Also
shown is the power delivered by Galactic supernovae,
\begin{equation}
\dot{W}_{SN} = E_{SN} \times \dot{SN} = 1.5 \times 10^{42} {\rm ~erg~s^{-1}},
\end{equation}
where $E_{SN}$=1.5$\times$10$^{51}$ erg is the approximate total
ejecta kinetic energy of a supernova (Woosley \& Weaver 1995).
LECRs with CRS$_{\rm metal}$ composition are the most efficient
soft X-ray emitters, because they produce intense line emission
from electron capture and excitation in 1-10 MeV/nucleon ions. We
see that for this composition and 3$<$$s$$<$3.5, the required LECR
power amounts to about 30\% of the available power in supernova
ejecta, allowing a reasonable LECR acceleration efficiency. We
note that the required acceleration is very modest, since the bulk
of the X-rays are produced by nuclei below about 10 MeV/nucleon,
and indeed the bulk of the LECR power also resides in such very
low energy particles. It is thus possible that the hard component
observed with {\it ASCA} is due to nonthermal ions. Of the three
assumed compositions, on grounds of energetics, CRS$_{\rm metal}$
appears to be the most promising.

\begin{figure}
\plotone{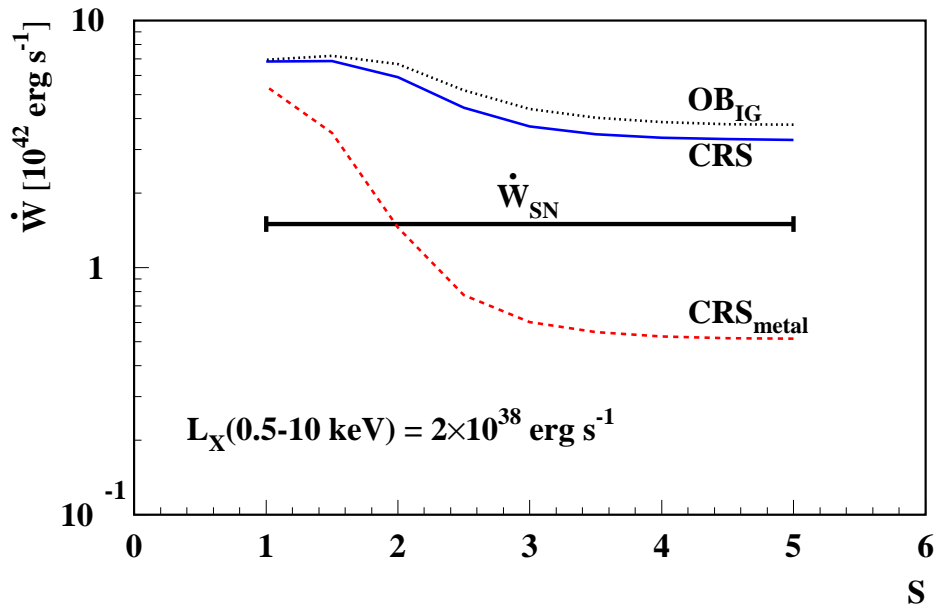} \caption{Power in LECRs required to produce
the estimated luminosity of the hard component detected with {\it
ASCA} in the 0.5-10 keV energy range, 2$\times$10$^{38}$ erg
s$^{-1}$ (Kaneda et al. 1997). For $s$$>$3, 90\% of the calculated
power resides in $<$10 MeV/nucleon particles. Also shown is the
power in supernova ejecta (eq. 3).}
\end{figure}

\begin{figure}
\plotone{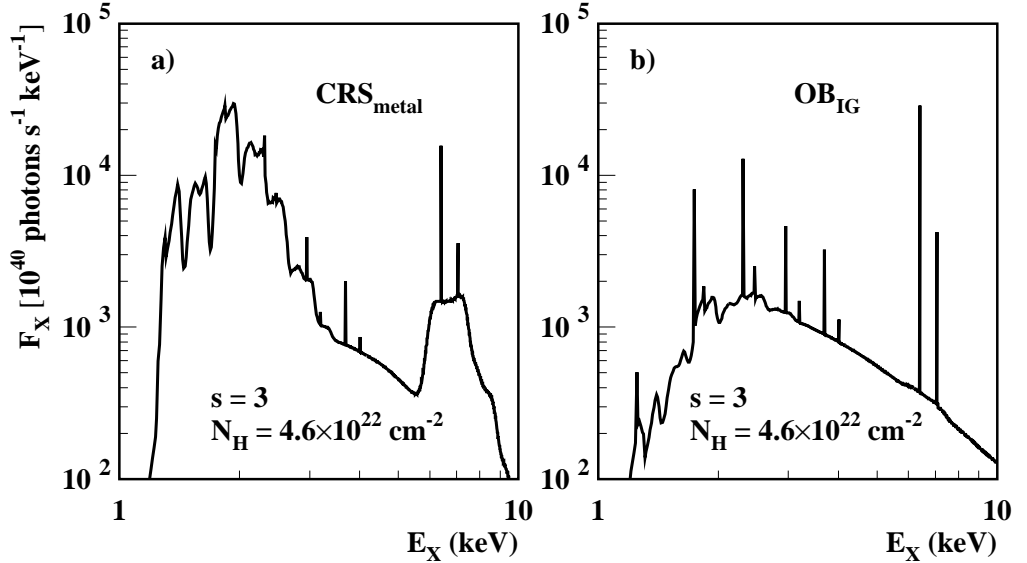}
\caption{Nonthermal X-ray emissions produced by LECRs with source spectral
index of 3 (eq.~1) interacting in a neutral ambient medium. (a)
CRS$_{\rm metal}$ composition. (b) OB$_{\rm IG}$ composition. Photoelectric
absorption is taken into account with a H column density of 4.6$\times$10$^{22}$
cm$^{-2}$. The calculations are normalized to a Be production rate of
3$\times$10$^{39}$ atoms s$^{-1}$.}
\end{figure}

Calculated nonthermal X-ray emissions are shown in Figures 3a and
b for the CRS$_{\rm metal}$ and OB$_{\rm IG}$ compositions,
respectively. We took into account photoelectric absorption with
N$_{\rm H}$=4.6$\times$10$^{22}$ cm$^{-2}$, the absorbing H column
density of the hard component in the {\it ASCA} energy band
(Kaneda et al. 1997). The narrow lines are due to K-shell vacancy
production in the ambient atoms by the fast ions (Tatischeff et
al. 1998). The broad line features, prominent in Fig. 3a, are due
to atomic de-excitations in the fast ions following charge
exchange (i.e. electron capture) and atomic excitation. The
detection of these broad lines would constitute an unequivocal
signature of LECRs in the Galaxy. Thus, the broad feature between
5.5 and 9 keV in Fig. 3a would trace the interactions of $\sim$10
MeV/nucleon Fe (Tatischeff et al. 1998). However, this excess is
nearly absent in Fig. 3b. We see that for the OB$_{\rm IG}$
composition, which is more impoverished in Mg, Si, S and Fe, only
the narrow lines from de-excitations in the ambient atoms could be
observed.

These narrow lines, due to K-shell vacancy production by ion
impacts, can be distinguished from the X-ray lines produced in a hot 
ionization equilibrium plasma, because their line energies are
different, e.g. the Fe K$\alpha$ line is at 6.40 keV following proton 
impact in a neutral medium while it is at 6.97 and 6.70 keV for H- and 
He-like Fe in a 7 keV plasma. However, Kaneda et al (1997) have shown that
the hard component detected with {\it ASCA} cannot be explained by a hot
plasma at ionization equilibrium, because the centroid energy of the
observed Fe line is lower than 6.70 keV. More detailed spectral analyses of
the {\it ASCA} data could provide further constrains on the possible 
contribution of nonthermal X-ray line emission. 
A fundamental problem is to distinguish the lines
produced by ion impacts from fluorescent lines, as the K-shell
vacancies created by protons and X-rays lead to line emission at
the same energies. However, for heavy ion collisions, the ion
impact lines could be shifted by several tens of electron-volts,
significantly broadened and splitted up into several components,
owing to multiple simultaneous ionizations (Garcia, Fortner \&
Kavanagh 1973). For example, the Fe K$\alpha$ line produced by 1.9
MeV/nucleon O impacts is blueshifted by $\sim$50 eV in comparison
with that produced by proton impacts, and has a FWHM of $\sim$100
eV (Garcia et al. 1973, figure 3.55). It may thus be possible with
future fine spectroscopic analyses to obtain the necessary
signatures to establish the existence of narrow lines produced by
LECRs.

\begin{figure}
\plotone{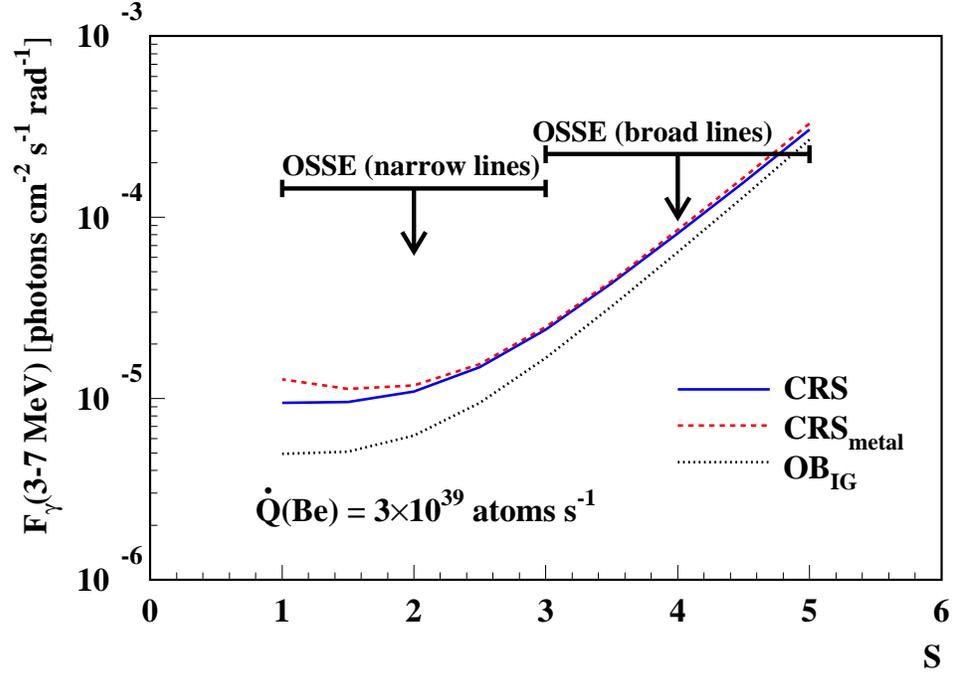} \caption{Predicted nuclear gamma-ray line
fluxes from the central radian of the Galaxy, assuming that that
all the current era Be production is due to LECR interactions,
with a Be production rate of 3$\times$10$^{39}$ atoms s$^{-1}$.
Also shown are the OSSE upper limits for broad and narrow line
emissions (Harris et al. 1996).}
\end{figure}

Finally, we calculated the 3-7 MeV nuclear gamma-ray line emission
that would accompany both Be and X-ray productions. The results
are shown in Figure 4. The predicted fluxes are compared with the
upper limits obtained with OSSE for both broad and narrow line
emissions from the central radian of the Galaxy (Harris et al.
1996). We used the same spatial model as Ramaty et al. (1997):
\begin{equation}
F_\gamma (3-7 \rm ~MeV) = \zeta 10^{-46} \dot{Q}_\gamma(3-7 \rm ~MeV)
{\rm ~~~(photons~cm^{-2}~s^{-1}~rad^{-1})}
\end{equation}
with $\zeta$=1. The total gamma-ray production rate
$\dot{Q}_\gamma(3-7 \rm ~MeV)$ is normalized to the Be production of
3$\times$10$^{39}$ atoms s$^{-1}$ (eq. 2). We see that for $s$$<$3.5, which
is the allowed range of spectral index in order to not overproduce the
Galactic ridge soft X-ray emission (Fig. 1b), the calculated gamma-ray
line fluxes are lower than 5$\times$10$^{-5}$ photons
cm$^{-2}$ s$^{-1}$ rad$^{-1}$. Similar results were obtained by Ramaty et al.
(1997) using different source spectra. Unfortunately, this diffuse
3-7 MeV emission should be difficult to detect with the {\it INTEGRAL}
spectrometer (Sch\"onfelder, this conference).

\section{Discussion}

Soft X-ray observations constitute an alternative and promising
way of tracing LECRs in the Galaxy. We have shown that the
distinct Galaxy wide LECR component, which has been proposed to
explain the recent Be abundance observations at low metallicities,
could account for the hard component of the GRXE in the 0.5-10 keV
energy domain, provided that the LECR spectrum extends as an
unbroken power law down to about 1 MeV/nucleon, implying that the
power in the accelerated particles is about 10$^{42}$ erg
s$^{-1}$. On the other hand, as the nonthermal X-rays in this
energy range are essentially produced below the thresholds of the
Be-producing cross sections, their detection would not ipso facto
imply a LECR origin for the spallogenic light elements.

However, as the expected nonthermal X-ray production is probably
not the dominant emission in the 10-60 keV energy range, it may
not be the main emission in the 0.5-10 keV energy range, as well.
A large population of low energy electrons, which could result
from efficient reacceleration of cosmic rays by interstellar
plasma turbulences (Schlickeiser 1997) may also contribute to the
Galactic ridge soft X-ray emission. In any case, it is important
to emphasize that the detection of K lines with {\it ASCA} does
not prove that the 0.5-10 keV GXRE is of thermal origin, since
interactions of accelerated ions could also produce intense K line
emission. In particular, the very low scale height of the hard
component detected with {\it ASCA} may be better understood if the
emission is of nonthermal origin. As suggested by Parizot (1998),
LECRs could be efficiently accelerated in superbubbles associated
with giant molecular cloud OB associations. In this model, we
would expect the nonthermal X-rays to be essentially produced at
dense molecular cloud boundaries, and thus the Galactic X-ray
emission to be correlated with the thin CO emission, which traces
the molecular hydrogen.

There are now evidences from both {\it ASCA} and RXTE surveys that the
diffuse X-ray emission from the Galactic plane has multiple origins. More
detailed spectral analyses of {\it ASCA} and future XMM data are required to
constrain a possible contribution of nonthermal ion interactions to the soft
X-ray emission and thus the eventual existence of a distinct low energy
cosmic ray component in the Galaxy.

\acknowledgments

We acknowledge B. Kozlovsky for useful discussions and J. Kiener for the 
careful reading of the manuscript.

\end{document}